\documentstyle[12pt]{article}
\global\parskip 6pt

\normalsize

\def\br{\begin{thebibliography}{abc}}
\def\er{\end{thebibliography}}

\def\be{\begin{equation}}
\def\ee{\end{equation}}
\def\bea{\begin{eqnarray}}
\def\eea{\end{eqnarray}}

\begin{document}
\begin{flushright}
\hfill{SINP/TNP/2009/27}\\
\end{flushright}
\vspace*{1cm}
\thispagestyle{empty}
\centerline{\bf Geometric Finiteness, Holography and Quasinormal Modes for the Warped ${\rm AdS}_3$ Black Hole }
\smallskip
\begin{center}
Kumar S. Gupta$^a$\footnote{Email: kumars.gupta@saha.ac.in}, E. Harikumar$^b$\footnote{Email: harisp@uohyd.ernet.in}, Siddhartha Sen$^{c,d}$\footnote{Email: sen@maths.ucd.ie} and M. Sivakumar$^b$\footnote{Email: mssp@uohyd.ernet.in}.

{\scriptsize {
$^a$ {\it Theory Division, Saha Institute of Nuclear Physics, 1/AF Bidhannagar, Calcutta 700064, India}.\\
$^b$ {\it School of Physics, University of Hyderabad, Hyderabad 500046, India}.\\
$^c$ {\it School of Mathematical Sciences, UCD, Belfield, Dublin 4, Ireland}.\\
$^d$ {\it Department of Theoretical Physics, Indian Association for the Cultivation of Science, Calcutta  700032, India}.\\ }}

\end{center}
\vskip.5cm

\begin{abstract}

We show that there exists a precise kinematical notion of holography for the Euclidean warped $AdS_3$ black hole. This follows from the fact that the Euclidean warped $AdS_3$ black hole spacetime is a geometrically finite hyperbolic manifold. For such manifolds a theorem of Sullivan provides a one-to-one correspondence between the hyperbolic structure in the bulk and the conformal structure of its boundary. Using this theorem we obtain the holographic quasinormal modes for the warped $AdS_3$ black hole.
\end{abstract}
\vspace*{.3cm}
\begin{center}
December 2009
\end{center}
\vspace*{1.0cm}

\section{Introduction}

BTZ black hole \cite{btz,btz1} provides an example of the AdS/CFT correspondence \cite{malda,sen1,danny1}. It has been suggested that there is a unique CFT associated with the BTZ black hole, which provides a toy model of quantum gravity \cite{witten1}. Recently a warped $AdS_3$ black hole has been found in topologically massive gravity \cite{deser1,deser2,strom1,carlip1}, where it was conjectured to be the holographic dual of a CFT \cite{strom}. 

For the Euclidean BTZ black hole, a mathematically precise kinematical description of holography was obtained \cite{sen1,sen2} using the idea of geometric finiteness \cite{sull,sull1}. The demonstration of the holography for the Euclidean BTZ was based on a mathematical theorem due to Sullivan, which states that for a geometrically finite manifold with boundary, there is a one-to-one correspondence between the hyperbolic structure of the interior and the conformal structure of the boundary \cite{sull,sull1}. The Sullivan's theorem subsequently led to a geometric description \cite{kumar1} of the quasinormal modes \cite{mann,nonqnm1} for the BTZ black hole, which also provides an evidence for the AdS/CFT correspondence for the BTZ \cite{danny1}.

In this Letter we establish that the warped Euclidean $AdS^3$ black hole is a geometrically finite hyperbolic manifold. From Sullivan's theorem, this implies the existence of a precise holographic correspondence between three dimensional hyperbolic structure of the warped $AdS^3$ black hole and the two dimensional conformal structure of its boundary, which is a warped two torus. This leads to relation between the monodromies of the solutions of the wave equations around the inner and outer horizons of the warped $AdS_3$ black hole. These monodromy relations are satisfied by a quantized set of frequencies. We call these the holographic quasinormal modes. Unlike the BTZ case \cite{nonqnm1,kumar1}, here the frequencies obtained from the monodromy relations are closely related but not identical to the quasinormal frequencies obtained using Dirichlet boundary condition \cite{oh}.  Quasinormal modes of massive scalar field for the warped $AdS_3$ black hole were obtained in \cite{wen,chen}.

This Letter is organized as follows. In Section 2 we show that the Euclidean warped $AdS^3$ black hole is a geometrically finite hyperbolic manifold, which implies a precise kinematical notion of holography following from the Sullivan's theorem. In Section 3, using Sullivan's theorem for the Euclidean warped $AdS^3$ black hole, we obtain geometric relations between the monodromies of the scalar fields around the inner and outer horizons of the black hole. We also find a set of quantized frequencies as solutions to these monodromy equations. Section 4 concludes the paper with some discussions.

\section{Holography for the Euclidean warped $AdS^3$ black hole}

The action for the 3D topologically massive gravity (TMG) with a negative cosmological constant $\Lambda = - \frac{1}{l^2}$, defined on a 3 manifold $M$ is given by \cite{deser1,deser2}
\be \label{action}
S_{TMG} = \frac{1}{16\pi G} \int_M d^3 x \sqrt{-g} \left [ (R + \frac{2}{l^2} )+ \frac{1}{\mu} S_{CS} \right ],
\ee
where $G$ is the Newton's constant and $\mu$ is the coupling of the gravitational Chern-Simons action. This model describes the a quantum theory of chiral gravity when $\mu l = 1$ \cite{strom1}. For other values of $\mu l \neq 1$ and for $\mu > 0$, although the $AdS_3$ vacua are unstable, the theory does admit stable warped $AdS_3$ solutions. It is useful to label the warped $AdS_3$ solutions of the 3D TMG in terms of a dimensionless constant $\nu = \frac{\mu l}{3}$. For $\nu^2 > 1$, this system admits black hole solutions which are free of closed timelike curves \cite{clement1,clement2}. In terms of 
Schwarzschild type coordinates ($r, t, \phi$), the spacelike stretched black hole metric is given by
\be \label{bhmetric}
ds^2 = -N^2 dt^2 + l^2 R^2 (d \phi + N^\phi dt)^2 + \frac{l^4 dr^2}{4 R^2 N^2},
\ee
where
\bea
R^2 &=& \frac{r}{4} \left [ 3 (\nu^2 - 1) r + (\nu^2 +3)(r_+ + r_-) - 4 \nu \sqrt{r_+ r_- (\nu^2 + 3)} \right ] , \\
N^2 &=& \frac{l^2 (\nu^2 + 3) (r - r_+) (r - r_-)}{4 R^2} , \\
N^\phi &=& \frac{2 \nu r - \sqrt{r_+ r_- (\nu^2 + 3)}}{2 R^2}.
\eea
The black hole radii are denoted by $r_+$ and $r_-$. The metric (\ref{bhmetric}) reduces to that for the BTZ black hole in a rotating frame when $\nu^2 = 1$ \cite{strom}.

The spacelike stretched ($\nu^2>1$) black hole described above admits a geometric construction \cite{strom}, very similar to that for the BTZ \cite{btz1}. In order to see this, first note that $AdS_3$ can be described as a fibration of $R$ over $AdS_2$. For such a spacelike fibration, the metric is given by
\be \label{spacelike}
ds^2 = \frac{l^2}{4} \left [ -d \tau^2 + d\alpha^2 + du^2 + 2 \sinh\alpha d\beta d\tau \right ],
\ee
where the coordinates $\alpha, \beta, \tau $ range from $-\infty$ to $\infty$. A warped spacelike $AdS_3$ metric can then be given by
\be \label{wsads3}
ds^2 = \frac{l^2}{\nu^2 + 3} \left [ - \cosh^2 \alpha d\tau^2 + d \alpha^2 + \frac{4 \nu^2}{\nu^2 + 3} (d \beta + \sinh \alpha d \tau)^2 \right ].
\ee
The key observation now is that the metric (\ref{bhmetric}) for the spacelike stretched warped $AdS_3$ black hole can be locally mapped to the metric (\ref{wsads3}) for the warped $AdS_3$ space using local coordinate transformations \cite{strom}. These two spaces therefore differ only in their global properties and both have constant negative curvature. Moreover, the
spacelike stretched warped $AdS_3$ black hole can be obtained from the warped $AdS_3$ space by quotienting the latter by a discrete subgroup of its isometry group \cite{strom}. Such an identification ensures the periodicity in the $\phi$ direction. The form of the identification made is given by
\be \label{group}
\Gamma = \{ \gamma^n \equiv e^{2 n \pi \xi},~ n \in Z \},
\ee 
where,
\bea \label{kill}
 \xi &=&  \frac{\nu^2 + 3}{8 \pi} \left [   A ~ \partial_\beta 
 +  B ~ ( \cos \tau ~ \tanh \alpha ~\partial_\tau + \sin \tau ~ \partial_\alpha + \frac{\cos \tau}{\cosh \alpha}~ \partial_\beta ) \right ], \\
A &=& r_+ + r_- - \frac{\sqrt{(\nu^2 + 3) r_+ r_-}}{\nu},   \\
B &=&  r_+ - r_-.
\eea
The set $\Gamma$ defines a discrete one parameter subgroup of the isometries of the warped $AdS_3$. The spacelike stretched warped $AdS_3$ black hole is obtained from the warped $AdS_3$ space by identifying points under the action of $\gamma^n,~ \gamma = e^{2 \pi \xi}$, $n \in Z $. 
Thus a point of the warped $AdS_3$ space with coordinates $(\alpha, \beta, \tau)$ is identified with
\be \label{ident}
(\alpha^{\prime},~\beta^{\prime},~\tau^{\prime}) \equiv  \gamma^n (\alpha,~\beta,~\tau) \sim (\alpha,~\beta,~\tau), ~~~ n \in Z.
\ee
Below consider the Euclidean continuation of the above spacetime. Since the hyperbolic character of the space is not  affected by the warping, the Euclidean continuation of the warped $AdS_3$ has a 3D hyperbolic structure as well.

We now show that the space obtained by this identification is a geometrically finite hyperbolic manifold. Let $\xi_E$ denote the Euclidean continuation of the Killing vector $\xi$ and let the corresponding one parameter subgroup be denoted by 
\be \label{gammae}
\Gamma_E = \{ \gamma_E^n \equiv e^{2 n \pi \xi_E},~ n \in Z \}.
\ee
$\Gamma_E$ is a discrete one parameter subgroup of the isometries of the Euclidean warped $AdS_3$ and is therefore a cyclic Kleinian group \cite{rat} with elements $ \gamma_E^n,~n \in Z$. Under the action of $\gamma_E^n$, an arbitrary point $(\alpha_E, \beta_E, \tau_E)$ of the warped Euclidean $AdS_3$ manifold goes to
\be \label{simi}
\gamma_E^n~: ~(\alpha_E, \beta_E, \tau_E) \rightarrow 
 e^{2 n \pi  \xi_E } (\alpha_E,~\beta_E,~\tau_E),
\ee
where the subscript $E$ denotes the Euclidean continuation and specifically we have  $\tau_E = i \tau$. The Killing vector $\xi$ in (\ref{kill}) and the corresponding Euclidean version $\xi_E$ are linear in the partial derivatives of the coordinates. In terms of $ \tilde{\alpha}= \frac{\alpha_E}{n}, ~\tilde{\beta} =\frac{\beta_E}{n},~ \tilde{\gamma}= \frac{\gamma_E}{n}$, we can rewrite (\ref{simi}) as 
\be \label{multi}
(\alpha_E, \beta_E, \tau_E) \rightarrow 
e^{2 \pi  \xi_E (\partial_{\tilde{\alpha}}, \partial_{\tilde{\beta}}, \partial_{\tilde{\gamma}})} (\tilde{\alpha} n, \tilde{\beta} n, \tilde{\gamma} n).
\ee
Therefore, in the limit $n \rightarrow \pm \infty $, an arbitrary point  $(\alpha_E,~\beta_E,~\tau_E)$ of the warped Euclidean $ADS_3$ goes to $(\pm \infty, \pm \infty, \pm \infty )$ respectively. To see the nature of this limit set, note that the metric for the Euclidean warped $ADS_3$ is given by 
\be \label{nm}
ds_E^2 = \frac{l^2}{\nu^2 + 3} \left [ - \cosh^2 \alpha_E d\tau_E^2 + d \alpha_E^2 + \frac{4 \nu^2}{\nu^2 + 3} (d \beta_E + \sinh \alpha_E d \tau_E)^2 \right ].
\ee
It is therefore clear that as $\alpha_E,~\beta_E,~\tau_E \rightarrow \pm \infty$, the metric (\ref{nm}) for the Euclidean warped $ADS_3$ collapses to a one dimensional metric corresponding to the $\tau_E$ direction only. From the above analysis we can conclude that the limit set of $\Gamma_E$ consists of only two points, given by $\infty$ and $-\infty$. Moreover, the limit set is independent of the chosen starting point. Since the limit set of $\Gamma_E$ is finite and consists of only two points, the Kleinian group $\Gamma_E$ elementary, and hence geometrically finite \cite{maskit}. It therefore follow that the spacelike stretched Euclidean $AdS_3$ black hole is a geometrically finite Kleinian manifold. The boundary of this space is topologically equivalent to a two torus.

From the above analysis it follows that the Euclidean warped $ADS_3$ black hole spacetime satisfies the conditions for the validity of the Sullivan's theorem.This implies that there exists a one-to-one correspondence between the hyperbolic structure of the interior of the Euclidean warped $ADS_3$ black hole and the conformal structure of its boundary, the latter being labelled by the Teichmuller parameter of a two torus. We therefore conclude that there exists a precise mathematical kinematical notion of holography for the Euclidean warped $ADS_3$ black hole. 

\section{Holographic quasinormal modes}

For the BTZ black hole, quasinormal modes for a scalar field can be obtained either by imposing Dirichlet boundary conditions at infinity or by imposing certain relations between monodromies of the solutions of Klein-Gordon equations around the black hole horizons \cite{nonqnm1}. The notion of holography, via the Sullivan's theorem, provides a geometrical basis for these monodromy relations \cite{kumar1}. In this Section we shall obtain similar monodromy relations for the warped $ADS_3$ black hole and find the frequencies that would satisfy these relations.

The boundary of the warped $ADS_3$ black hole is topologically equivalent to a two torus $T^2$ and the corresponding Teichmuller space is given by a fundamental region of the complex variable $\tau$. Two Teichmuller parameters $\tau$ and $\tau^\prime$ are equivalent if
\be
\tau^\prime =\frac{a\tau +b}{c \tau +d},\quad   ad-bc=0
\ee
and $a,b,c,d \in Z$.
The transformation is the action of the modular group on $\tau$ generated by 
\be
S: \tau \to -\frac{1}{\tau};\quad
T:\tau\to \tau +1
\ee
For the warped $ADS_3$ black hole, the inner and outer radii can be interchanged by a suitable combination of these generators, which we denote by $Q \in SL(2,Z)$. 

We proceed to determine the monodromies of the solutions of the massless scalar field equation in the background of warped $ADS_3$ black hole \cite{oh} about the horizons $r=r_\pm$. The wave equation is given by
\be 
\nabla_L^2 \chi=0
\ee
Setting $\chi = \Psi(r)\exp \{i\omega t +i \mu \theta \} ,\Psi(r)$ was found to be 
\begin{eqnarray}
\Psi(r) &=&C_1 (r-r_-) ^{-A} (r-r_+)^B F[{\cal A}_-, {\cal B}_- ,{\cal C}_-;\frac{r_+-r}{r_+-r_-} ]\nonumber\\
&+&C_2 (r-r_-) ^{-A} (r-r_+)^{-B} F[{\cal A}_+, {\cal B}_+,{\cal C}_+;\frac{r_+-r}{r_+-r_-} ],\label{wave}
\end{eqnarray}
where
\begin{eqnarray}
& A=\frac{\sqrt{\alpha r_{-}^2+\beta r_-+\gamma}}{r_+-r_-},~~
B=\frac{\sqrt{\alpha r_{+}^2+\beta r_+ +\gamma}}{r_+-r_-},&\\
&{\cal A}_\pm={\mp B}-A+\frac{1}{2}(1\pm\sqrt{1+4\alpha}),&\\
&{\cal B}_\pm={\cal A}_\pm \mp \sqrt{1+4\alpha},~~{\cal C}_\pm= 1\pm 2B.&
\end{eqnarray}
In the above, we have used,
\begin{eqnarray}
&\alpha\equiv -3\omega^2(\nu^2-1)(\nu^2+3)^{-2},&\\
&\beta\equiv -[\omega^2(\nu^2+3)(r_++r_-)-4\nu(\omega^2\sqrt{r_+r_-(\nu^2+3)}-2\mu\omega)](\nu^2+3)^{-2},&\\&\gamma\equiv-4[\mu^2-\omega\mu\sqrt{r_+r_-(\nu^2+3)}~](\nu^2+3)^{-2}&
\end{eqnarray}
Let $P(r_+)$ denote the monodromy of the solution of the Klein-Gordon equation around the outer horizon $r_+$. For this purpose we consider only that part of the solution which is purely ingoing at $r = r_+$ \cite{nonqnm1}, which is obtained from Eqn. (\ref{wave}) with $C_1=0$ \cite{oh}, given by
\be
\Phi(r) = C_2 (r-r_-) ^{-A} (r-r_+)^{-B} F[{\cal A}_+, {\cal B}_+,{\cal C}_+;\frac{r_+-r}{r_+-r_-} ].\label{wavesoln}
\ee 
The monodromy depends on the metric of the warped $ADS_3$ black hole, or equivalently on the hyperbolic structure. On the other hand, the conformal structure of the boundary is defined by the Teichmuller parameter. As stated before, two Teichmuller parameters related by the action of the modular group are identified. Since Sullivan's theorem provides a one-to-one correspondence between the hyperbolic structure and the conformal structure, the monodromies related by the modular transformations should be identified as well. Therefore, under the action of $Q \in SL(2,Z)$, we have
\be 
QP(r_+)=P(r_+).
\ee 
Note that $Q$ is the transformation which interchanges $r_+$ with $r_-$. Therefore, we also have
\be
QP(r_+) = P(Qr_+) = P^{\pm}(r_-),
\ee \label{monocond}
where the $\pm$ sign correspond to the two analytic continuations of Eqn. (\ref{wavesoln}) to $r=r_-$. 

We first calculate the monodromy using the ingoing part of $\Phi(r)$ given in Eqn.(\ref{wavesoln}). The monodromy at $r=r_+$ can be read off easily from (\ref{wavesoln}) and is given by
\be
P(r_+) =  e^{-2\pi iB}. 
\ee \label{mono1}
To find the same at $r=r_-$, first we analytically continue Eqn.(\ref{wavesoln}) to $r\to r_-$. For this, we use the  transformation of the hypergeometric function \cite{MAIAS},
\begin{eqnarray}
F(a,b;c,z)&=&\frac{\Gamma(c)\Gamma(c-a-b)}{\Gamma(c-a)\Gamma(c-b)}F(a;b,a+b=c;1-z)\nonumber\\&+& (1-z)^{c-a-b} 
\frac{\Gamma(c)\Gamma(a+b-c)}{\Gamma(a)\Gamma(b)} F(c-a,c-b;c-a-b+1,1-z).
\end{eqnarray} \label{analcont}
Using Eqns. (\ref{wavesoln}) and (\ref{analcont}), we find that the monodromies at $r=r_-$ as
\begin{equation}
 P^+(r_-)= e^{-2\pi iA},~~~P^-(r_-)= e^{2\pi i(A+4B)}
\end{equation}
Finally, using the conditions on the monodromies following from the Sullivan's theorem given by $P(r_+) = P^{\pm}(r_-)$ and using Eqns. (21)-(26), we find the holographic quasinormal frequencies as 
\begin{eqnarray}
\omega_L&=& i\frac{(\nu^2+3)n}{2\nu},\nonumber\\
\omega_R&=& -\frac{(6\mu+\frac{in}{2}(\nu^2+3)(r_+-r_-)}{\nu(r_-+5r_+)-3\sqrt{r_+r_-(\nu^2+3)}}
~n\in Z
\end{eqnarray}
Since these frequencies are obtained using the Sullivan's theorem, we call these as the holographic QNM frequencies. The $\omega_L$ is consistent with that obtained in \cite{oh} whereas the expression for $\omega_R$ is different from that obtained using the Dirichlet condition. We make further comments about this in the concluding section.  

\section{Conclusions}

In this Letter we have shown that the warped Euclidean $ADS_3$ black hole is a geometrically finite manifold. For such manifolds, Sullivan's theorem provides a one-to-one correspondence between the hyperbolic structures of the interior and the conformal structures of the boundary. Thus the theorem provides a precise kinematical notion of holography (AdS/CFT correspondence) which was conjectured in \cite{strom}. 

The Sullivan's theorem also provides a relation between the monodromies of the solutions of the wave equation around the inner and outer horizons of the warped $ADS_3$ black hole. For the BTZ case, the solution of such monodromy relations leads to the QNM frequencies \cite{nonqnm1,kumar1}. For the warped $ADS_3$ black hole the monodromy relations are satisfied by a quantized set of frequencies. However, one set of these holographic QNM frequencies differ from the QNM obtained from the Dirichlet condition \cite{oh}. This is very different behaviour compared to the BTZ case, where the nonqnm frequencies obtained from the monodromy conditions agree with that obtained using the Dirichlet condition \cite{nonqnm1}. 

Quasinormal modes obtained using the Dirichlet boundary conditions can be shown using general considerations to determine the poles of the retarted Green's function of the two-point function corresponding to the operator in the dual CFT of the scalar field in the bulk, see \cite{berti} for a recent review. Since the holographic quasi-normal modes differ from that obtained by the Dirichelt boundary conditions, then it is clear they do not determine the poles of the Green's function. It will be certainly interesting to see what these modes determine for the dual CFT Green's function as the authors mention.

In order to investigate these issues, it would be important to analyze the dual CFT \cite{danny1} for the warped $AdS_3$ case to find the relaxation time scales directly, which will allow the comparison of the CFT results with those obtained here and with those obtained from the Dirichlet condition.

\bigskip
\noindent
{\bf Acknowledgments}

One of us (SS) thank School of Physics, University of Hyderabad, where a part of this work was done while visiting as Jawaharlal Nehru Chair Professor.

\bibliographystyle{unsrt}

\end{document}